\documentclass{article}

\usepackage{arxiv}

\usepackage[utf8]{inputenc} 
\usepackage[T1]{fontenc}    
\usepackage{hyperref}       
\usepackage{url}            
\usepackage{booktabs}       
\usepackage{amsfonts}       
\usepackage{nicefrac}       
\usepackage{microtype}      
\usepackage{lipsum}		
\usepackage{graphicx}
\usepackage{doi}
\usepackage{float}
\usepackage{listings} \usepackage{xcolor}

\usepackage{array}
\usepackage{makecell}
\usepackage{geometry}
\newcolumntype{C}[1]{>{\centering\arraybackslash}m{#1}}

\title{Decentralized Vulnerability Disclosure via Permissioned Blockchain: A Secure, Transparent Alternative to Centralized CVE Management}

\author{ \href{https://orcid.org/0000-0000-0000-0000}{\includegraphics[scale=0.06]{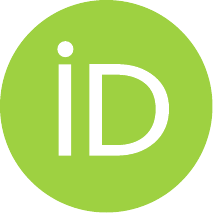}\hspace{1mm}Novruz Amirov}\\
	Institute of Informatics\\
	Istanbul Technical University\\
	Istanbul, Turkey \\
	\texttt{amirov20@itu.edu.tr} \\
	\And
	\href{https://orcid.org/0000-0000-0000-0000}{\includegraphics[scale=0.06]{orcid.pdf}\hspace{1mm}Kemal Bicakci} \\
	Institute of Informatics\\
	Istanbul Technical University\\
        Istanbul, Turkey \\
	\texttt{kemalbicakci0@itu.edu.tr} \\
}

\renewcommand{\shorttitle}{Decentralized CVE Publication via Permissioned Blockchain}

\hypersetup{
pdftitle={A template for the arxiv style},
pdfsubject={q-bio.NC, q-bio.QM},
pdfauthor={David S.~Hippocampus, Elias D.~Striatum},
pdfkeywords={First keyword, Second keyword, More},
}

\begin{document}
\maketitle

\begin{abstract}
This paper proposes a decentralized, blockchain-based system for the publication of Common Vulnerabilities and Exposures (CVEs), aiming to mitigate the limitations of the current centralized model primarily overseen by MITRE. The proposed architecture leverages a permissioned blockchain, wherein only authenticated CVE Numbering Authorities (CNAs) are authorized to submit entries. This ensures controlled write access while preserving public transparency. By incorporating smart contracts, the system supports key features such as embargoed disclosures and decentralized governance. We evaluate the proposed model in comparison with existing practices, highlighting its advantages in transparency, trust decentralization, and auditability. A prototype implementation using Hyperledger Fabric is presented to demonstrate the feasibility of the approach, along with a discussion of its implications for the future of vulnerability disclosure.
\end{abstract}


\section{Introduction}
\begin{figure}[H]
    \centering
    \includegraphics[width=0.9\linewidth]{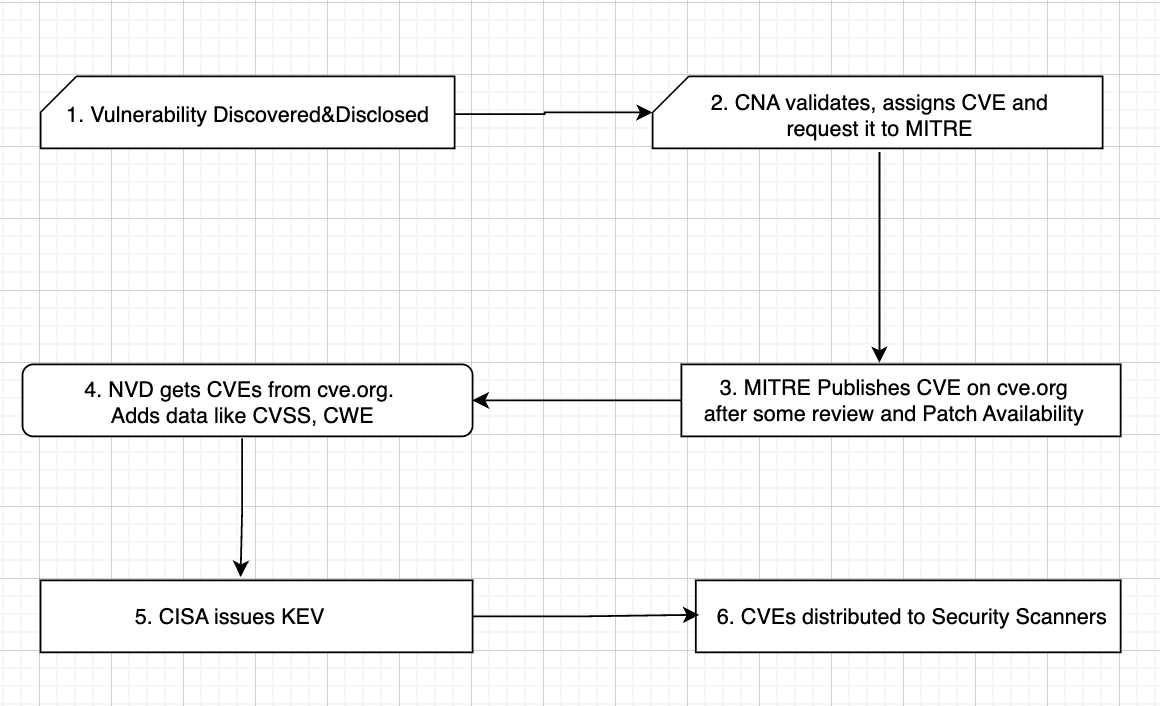}
    \caption{Current CVE Publication Process\cite{mitre-cve-program}}
    \label{cve-process}
\end{figure}

Vulnerability disclosure is a cornerstone of cybersecurity. The current ecosystem for disclosing security vulnerabilities, particularly via the CVE (Common Vulnerabilities and Exposures) system, is coordinated primarily by MITRE Corporation and relies on a centralized structure of CVE Numbering Authorities (CNAs). Currently, for new CVE-ID publication, first the vulnerability is discovered by known CNAs, the it is submmitted to MITRE where it is reviewed and if it is verified, the CVE is published on cve.org. The National Vulnerability Database, or NVD, gets CVE from the official website of Mitre, and adds data such as CVSS, CWE, and etc. Then, the CISA Known Exploited Vulnerabilities (KEV) catalog is updated for the according vulnerability. Finally, the security scanners is updated as the process is described on Figure \ref{cve-process}.

\subsection{Statistics concerning current process}
\par To better understand the current landscape of vulnerability disclosures, recent statistics from OpenCVE reveal insightful trends, mentioning scalability over the years \cite{current-statistics}.

\par
\begin{figure}[H]
    \centering
    \includegraphics[width=0.9\linewidth]{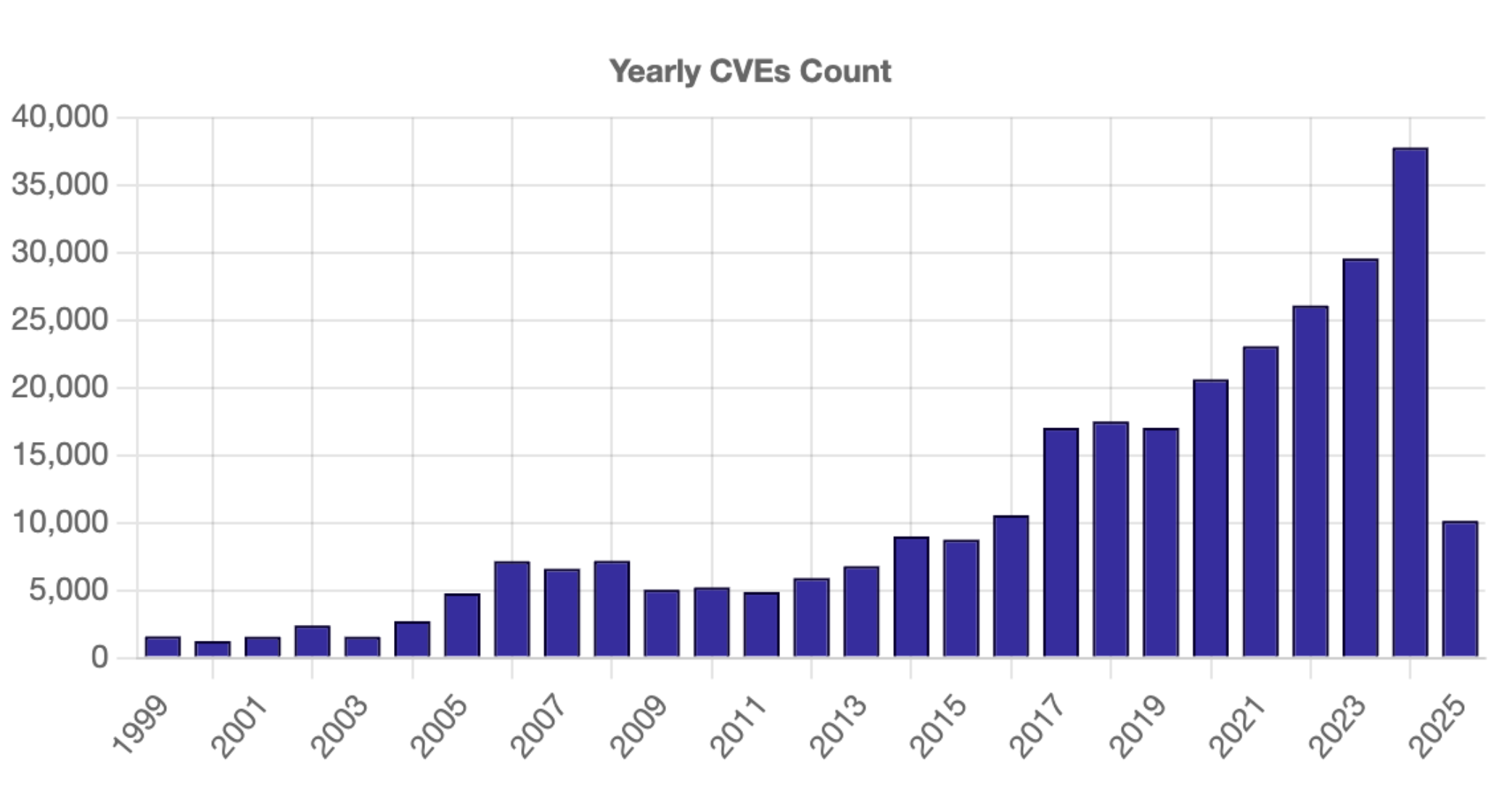}
    \caption{CVE statistics over the years since 1999.}
    \label{yearly-stats}
\end{figure}

 As of 2025, over 10,000 CVEs have been published inspite of the year not completed, reflecting a steady increase in vulnerability disclosures compared to previous periods, as shown in Figure \ref{yearly-stats}. 

\begin{figure}[H]
    \centering
    \begin{minipage}{0.48\textwidth}
        \centering
        \includegraphics[width=\linewidth]{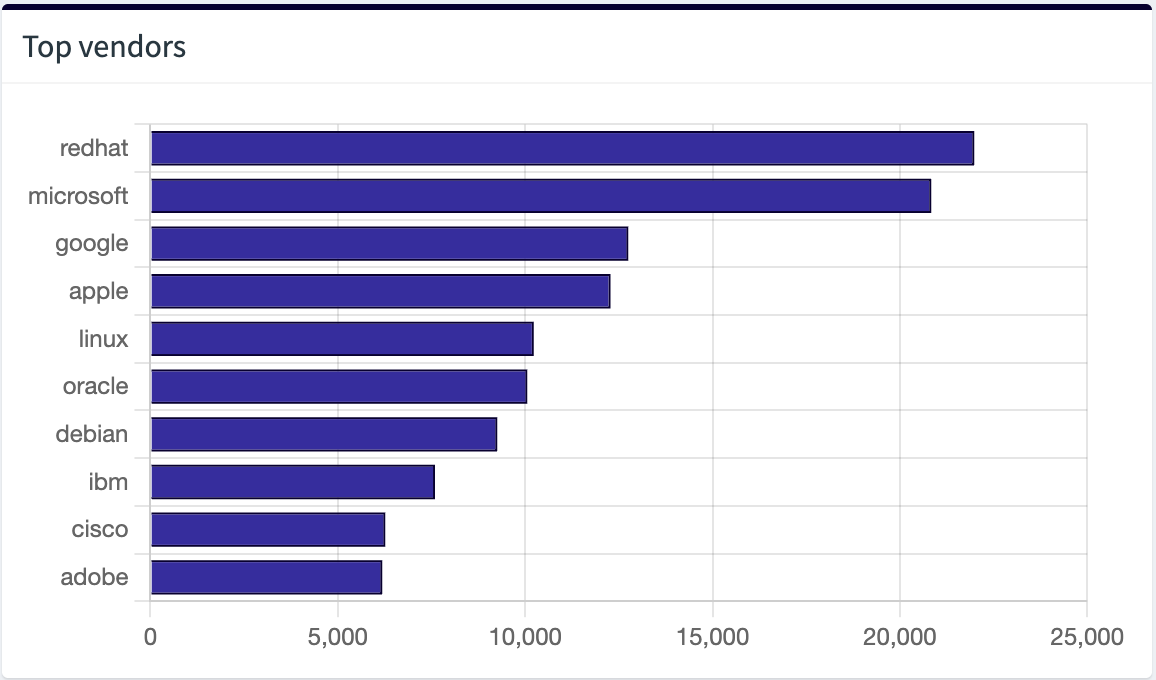}
        \caption{Top Vendors having CVEs identified}
        \label{top-vendors}
    \end{minipage}
    \hfill
    \begin{minipage}{0.48\textwidth}
        \centering
        \includegraphics[width=\linewidth]{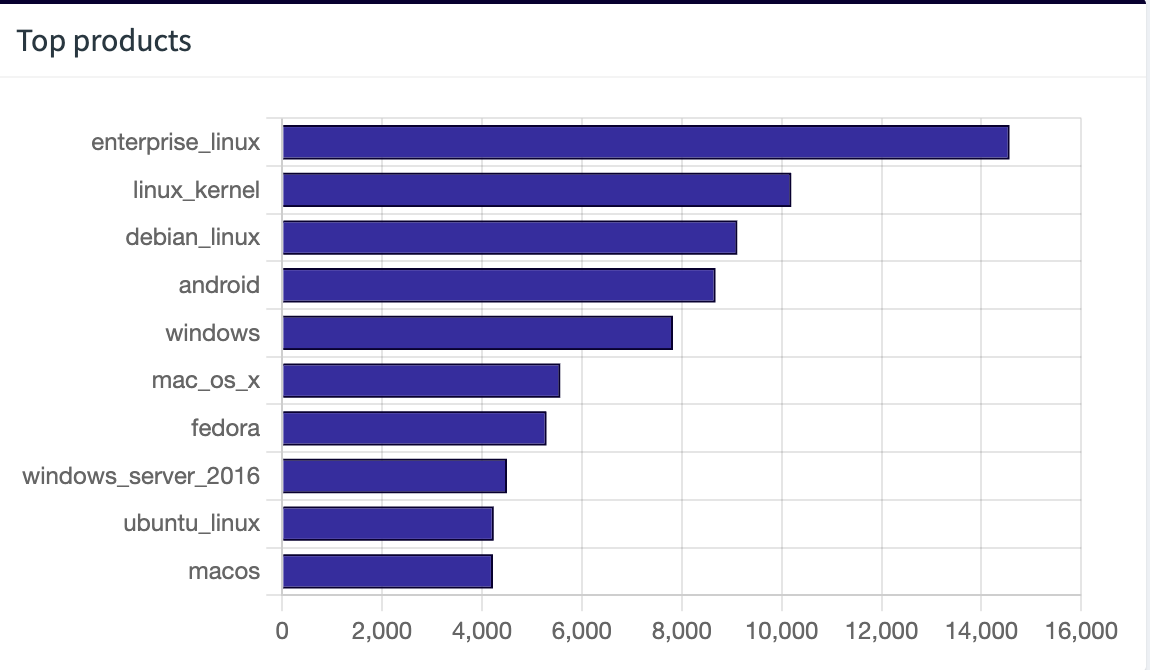}
        \caption{Top Products having CVEs identified}
        \label{top-products}
    \end{minipage}
\end{figure}

\par The top vendors associated with CVE publications include Microsoft, Google, and Oracle, as displayed in Figure \ref{top-vendors} while prominent products such as Windows 10, Chrome, and Android consistently appear among the most affected, as shown in Figure \ref{top-products}. These patterns highlight the growing complexity and scale of modern vulnerability management ecosystems

\subsection{Current Counting Issues on CVEs}
\par The process of managing and maintaining the accuracy of CVE entries is critical to ensuring the quality and reliability of the CVE system. Given the complexity and volume of vulnerability disclosures, errors are inevitable. The CVE program has thus defined formal mechanisms to correct issues, which are categorized into rejection, merging, splitting, disputing, and handling partial duplicates. \cite{current-issues}

\par CVE assignment errors commonly arise due to \textbf{insufficient information} which is a result of inadequate research into the releationship between codebases and products. Moreover, \textbf{inadequate coordination} where multiple CNAs assigns CVE IDs independently without synchronization is another cause. \textbf{Human Error} like typograpchical errors in vulnerability reports are also posisble to cause an issue. 
Corrections may be initiated by the CNA that assigned the CVE or a third-party entity providing additional information.

\par Correction process includes \textbf{Reject, Merge, Split, Dispute, and Partial Duplicate}.
\begin{itemize}
    \item \textbf{Reject:} A CVE ID may be rejected when it is determined that the assigned issue does not constitute a vulnerability, or when administrative errors occur, such as typographical mistakes or the withdrawal of a report. In this case, The CVE entry description is updated with the status \textbf{"REJECTED".} Moreover, an explanation for the rejection is provided. Finally, the rejected CVE remains visible but is annotated to avoid future confusion. 

    \item \textbf{Merge: }When multiple CVE IDs are assigned to the same vulnerability, a merge process is initiated. One CVE ID is selected as the canonical identifier based on: 1) The most commonly referenced identifier. 2) The identifier used by the most authoritative source (vendor > coordinator > researcher). 3) The earliest publicized identifier. 4) The identifier with the smallest numeric portion. 
    \par Other CVE IDs are updated with a "REJECTED" status, pointing to the selected CVE ID. This ensures consolidation and reduces redundancy in vulnerability tracking.

    \item \textbf{Split:} When a single CVE ID covers multiple independent vulnerabilities, a split operation is required. The original CVE ID is retained for the most prominent vulnerability, determined by: 1) Common association frequency. 2) Severity of risk (based on CVSS scores). 3) Breadth of affected versions. 4) Order of mention in the initial publication. 
    \par New CVE IDs are assigned for additional vulnerabilities. Cross-references are added among the affected CVE entries. This process ensures a one-to-one mapping between vulnerabilities and identifiers, in accordance with the CVE granularity principles.

    \item \textbf{Dispute:} A dispute arises when the validity of a vulnerability is contested. The CVE description is prefixed with \textbf{"DISPUTED"}. A NOTE is appended explaining the nature of the dispute, ideally supported by an external reference or quote. Disputes typically concern the interpretation of vulnerability conditions, rather than administrative or technical errors.
    Unlike rejections, disputes acknowledge the existence of conflicting perspectives rather than errors in assignment.

    \item \textbf{Partial Duplicate}: Occurs when two CVE IDs reference overlapping but not identical sets of affected software versions. The CVE ID that is most commonly referenced or issued by the most authoritative source is prioritized. The overlapping CVE is revised to cover only the non-overlapping versions. Mutual references are added, noting the relatedness of the entries. If necessary, additional criteria such as earliest publication or smallest numeric portion guide the resolution.
\end{itemize}

\subsection{What is proposed as a solution to current process?}
\par While this system has been instrumental in standardizing vulnerability reporting, it presents several risks due to its centralized nature—such as single points of failure, lack of global transparency, and governance limitations.

The increasing complexity and volume of security disclosures necessitate a more robust, transparent, and globally accessible infrastructure. In this work, we propose a blockchain-based architecture where CNAs serve as validator nodes within a permissioned system. Our design allows for transparent publication of verified vulnerabilities while maintaining strict submission permissions and enabling advanced features like embargo mechanisms and DAO-based governance.

This paper contributes:
\begin{enumerate}
    \item Why there is a need for blockchain? Which type of blockchain shoul it be used?
    \item A secure, permissioned blockchain architecture for CVE publication.
    \item Smart contract designs for enforcing CNA authentication, embargo timing, and governance.
    \item A prototype implementation using Hyperledger Fabric.
    \item A comparative analysis with centralized CVE disclosure models.
\end{enumerate}

\section{Blockchain for Decentralized Management of CVEs}
\par Blockchain is a decentralized, tamper-evident ledger technology that enables trustless collaboration and transparent data sharing across multiple parties. It has been widely explored in domains requiring verifiability, auditability, and secure information exchange. In the context of CVE publication, blockchain offers an opportunity to overcome the inherent risks of centralized control, such as the dependency on a single trusted entity like MITRE.

To assess whether blockchain is appropriate for decentralized CVE management, we adopt the decision framework from Wüst and Gervais \cite{first} as displayed in Figure \ref{decision-methodology}:

\begin{figure}[H]
    \centering
    \includegraphics[width=0.9\linewidth]{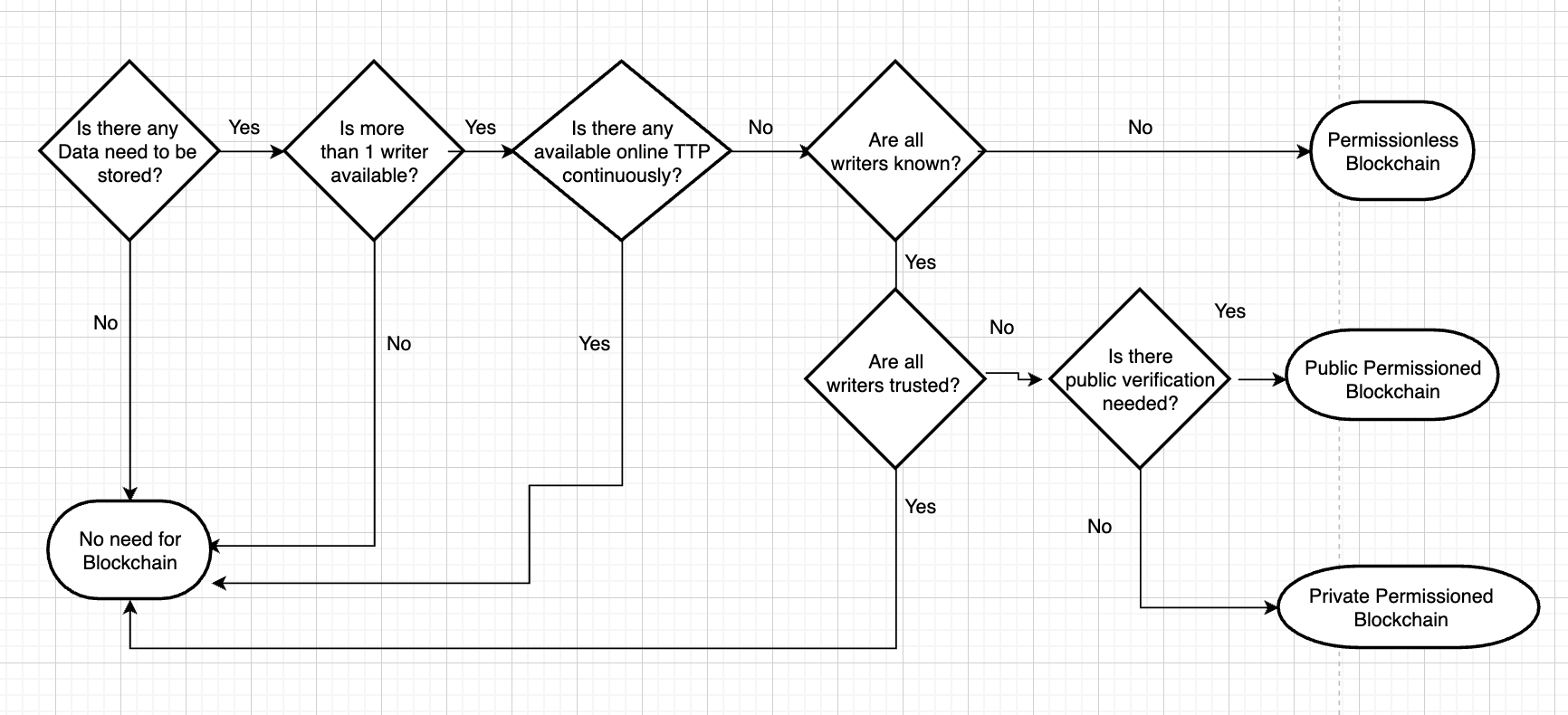}
    \caption{Flowchart to decide whether Blockchain should be used or not. If so, which type of it as a best solution.}
    \label{decision-methodology}
\end{figure}

\begin{enumerate}
    \item Do you need to store data? → Yes. Our system must store structured data including CVE identifiers, metadata (affected vendors, impact, CVSS score), disclosure dates, and update histories. This data needs to be publicly accessible and tamper-proof.
    \item Are there multiple writers? → Yes. Different CVE Numbering Authorities (CNAs) globally contribute vulnerability disclosures, making them independent writers.
    \item Can you use an always online Trusted Third Party (TTP)? → No. Reliance on MITRE has already demonstrated the fragility of centralized governance. For example, recent concerns about the potential withdrawal of funding from MITRE led to fears about the stability of the entire CVE system. A decentralized system eliminates this single point of failure.
    \item Are all writers known? → Yes. All CNAs are publicly listed in the CNA registry \cite{second}, enabling a permissioned environment where write privileges are limited to pre-approved entities.
    \item Are all writers trusted? → No. Despite being known, CNAs may vary in practices and reliability. Malicious or negligent behavior—such as submitting false disclosures or leaking embargoed vulnerabilities—necessitates cryptographic guarantees and programmatic governance.
    \item Is public verifiability required? → Yes. Public trust in CVEs depends on the ability of researchers, vendors, and organizations to independently verify their authenticity and integrity. Blockchain inherently provides this property through distributed consensus and audit logs.
\end{enumerate}

Given these criteria, the most appropriate architecture is a public permissioned blockchain. This design allows a restricted set of authorized CNAs to submit vulnerability data (permissioned writers), while allowing anyone to verify and audit the information (public readability). Hyperledger Fabric is chosen as our underlying technology due to its modular architecture, identity management, and support for chaincode (smart contracts).

\section{Permissioned Blockchain usage for CVEs}

\subsection{System Design and Architecture: Permissioned}
\par Our system leverages Hyperledger Fabric, a modular and extensible permissioned blockchain framework, to manage the publication and lifecycle of Common Vulnerabilities and Exposures (CVEs) in a decentralized yet controlled environment. Each CVE Numbering Authority (CNA) is represented as an independent organization within the Fabric network and operates its own peer node. A consortium of governance entities operates the ordering service to preserve consensus and ensure network integrity.

The architecture supports modular identity management, peer-to-peer transaction validation, and customizable endorsement policies. This setup ensures strict authentication of actors (CNA organizations) while maintaining publicly auditable records of CVEs. Our smart contract (chaincode) modules enforce data integrity, embargo policies, and role-based access control. The full source code for this implementation is publicly available at Github Repository \cite{github-repo}.

\subsection{CNA Authentication and Onboarding/Revoking}
\par Each CNA undergoes onboarding through a Certificate Authority (CA), which issues X.509 digital certificates managed by Hyperledger Fabric's Membership Service Provider (MSP). Onboarding requires passing KYC (know your customer) procedures and manual review by governance members, after which CNAs are granted permission to submit and update CVEs.

Revocation of CNA privileges is enforced through updates to the Certificate Revocation List (CRL) and smart contract-based governance checks. The system dynamically validates every transaction's submitter identity by matching the certificate to a list of authorized CNAs, maintaining a tamper-proof, accountable registry.

\begin{figure}[H]
    \centering
    \includegraphics[width=0.9\linewidth]{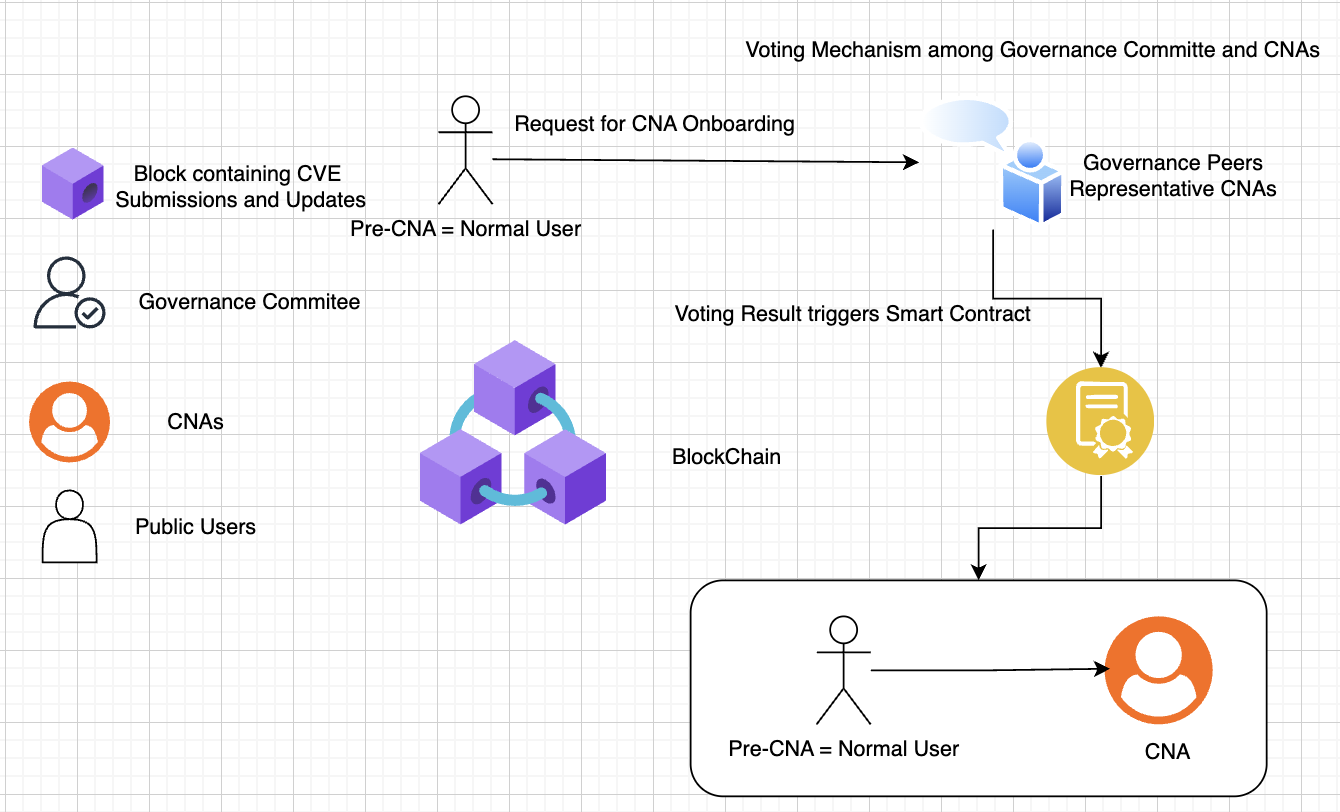}
    \caption{System Architecture and Roles}
    \label{system-architecture}
\end{figure}

\subsection{Smart Contract Design for CVE Lifecycle}
\par The chaincode is implemented in Go and defines the core logic for CVE management. It encapsulates:
\begin{enumerate}
    \item Schema enforcement (CVE ID, description, severity, product, embargo timestamp, etc.)
    \item Role-based access (submitter CNA vs governance)
    \item Embargo handling (delayed publication until a given timestamp)
    \item Event emission for off-chain indexing
    \item Lifecycle transitions (e.g., draft → published → archived)
\end{enumerate}

\par Key functions in the chaincode include \textbf{SubmitCVE} which verifies CNA identity, schema validity, and embargo constraints before registering the CVE, \textbf{UpdateCVEStatus} which allows only the original submitter to transition a CVE between statuses, \textbf{CheckEmbargoReleases} which periodically scans for and promotes embargoed CVEs once their release timestamp passes, \textbf{OnboardCNA and RevokeCNA} that is controlled by governance entities to manage the network's trust boundaries and etc.

Full smart contract source and deployment scripts are available in the chaincode/ directory of the github repository.


\begin{lstlisting}[language=Java, caption={Simplified Smart Contract Logic for CVE Management}, label={lst:smartcontract}] contract CVEChaincode {

struct CVE { string cveID; string description; string product; string version; string severity; string status; // draft, published, archived timestamp embargoUntil; string submitterCNA; }

function SubmitCVE(CVE input, callerID) { assert callerID IsIn authorizedCNAs; assert schemaIsValid(input); input.status <- (input.embargoUntil > now) ? "draft" : "published"; cveRegistry[input.cveID]<- input; emit Event("CVESubmitted", input.cveID); }

function UpdateCVEStatus(cveID, newStatus, callerID) { assert callerID == cveRegistry[cveID].submitterCNA; assert statusTransitionValid(oldStatus, newStatus); cveRegistry[cveID].status<- newStatus; }

function CheckEmbargoReleases() { foreach cve IsIn cveRegistry: if cve.status == "draft" AND cve.embargoUntil <= now: cve.status <- "published"; emit Event("EmbargoReleased", cve.cveID); }

function OnboardCNA(cnaID, certHash, callerID) { assert isGovernanceMember(callerID); assert verifyCert(cnaID, certHash); authorizedCNAs.add(cnaID); }

function RevokeCNA(cnaID, callerID) { assert isGovernanceMember(callerID); authorizedCNAs.remove(cnaID); } } \end{lstlisting}

\subsection{Data Flow}
\par The CVE data submission process is as follows. First, CNA prepares CVE metadata and signs it with their digital identity and submits the transaction to a peer node. Then, endorsement policy verifies the CNA identity and schema compliance. Endorsed transactions are ordered and committed to the ledger. Finally, Published CVEs become immediately auditable by the public.

\begin{figure}[H]
    \centering
    \includegraphics[width=0.8\linewidth]{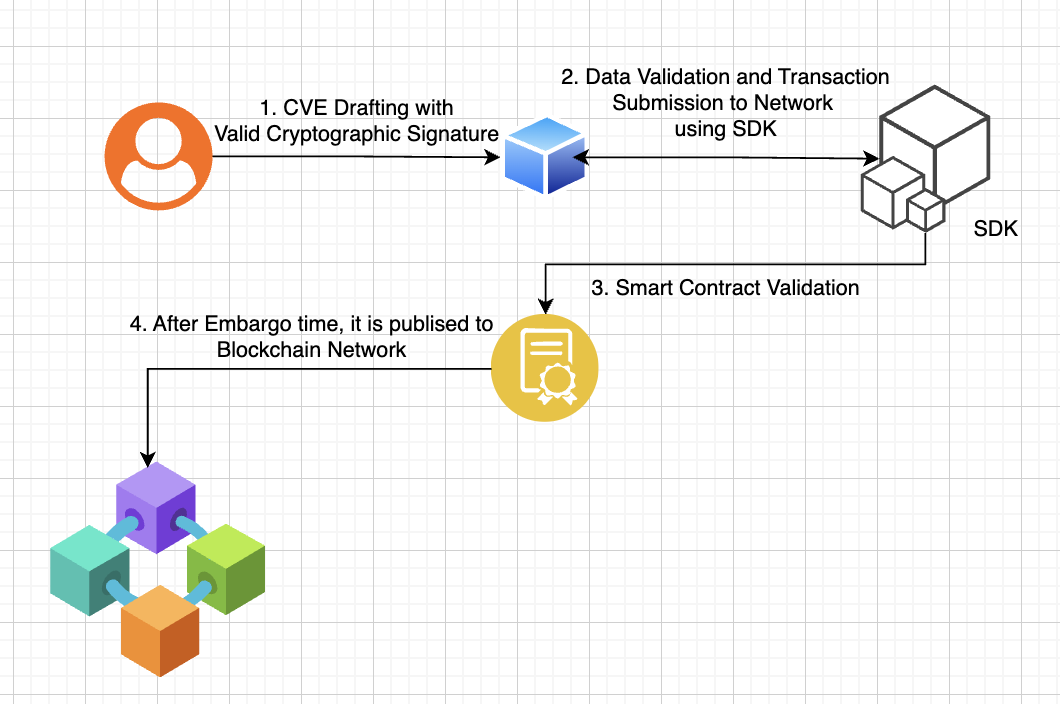}
    \caption{Data Flow of CVE Submission}
    \label{data-flow-cve}
\end{figure}

\subsection{Platform, Performance and Security Evaluation}
\par We use Hyperledger Fabric v2.4, deployed on a Kubernetes cluster with three organizations (CNAs), one ordering service, and a REST API gateway. Initial benchmarks using Hyperledger Caliper show throughput of 200 TPS with latency under 2 seconds for standard CVE transactions. For security, cryptographic identity management, endorsement policies, and tamper-evident logs ensure data authenticity. Chaincode enforces submission correctness and prevents unauthorized access.

\begin{table}[h!]
\centering
\begin{tabular}{|C{3.5cm}|C{6.5cm}|C{6.5cm}|}
\hline
\textbf{Evaluation Metric} & \textbf{Current Centralized Model (MITRE)} & \textbf{Proposed Blockchain-based Model} \\
\hline
Platform Architecture & Centralized database and management via MITRE and select CNAs & Distributed peer-to-peer network using Hyperledger Fabric \\
\hline
Write Access Control & Limited to MITRE & Enforced by smart contracts and digital certificates within permissioned blockchain \\
\hline
Data Availability & Single point of failure risk (e.g., if MITRE or CISA halts funding) & Highly available, replicated ledger across CNA peer nodes \\
\hline
Transparency & Limited public insight into CVE submission and update process & Full transaction history and public verifiability of all CVE states \\
\hline
Embargo Handling & Managed off-chain, opaque process & Smart contract-enforced embargo logic with cryptographic proofs \\
\hline
Performance (Latency) & Low submission throughput; manual verification and delays & ~2 seconds latency for endorsed transactions (based on Hyperledger Caliper tests) \\
\hline
Performance (Throughput) & Unknown, centralized bottleneck possible & Scalable with ~200 TPS in current benchmarks \\
\hline
Security (Authentication) & Manual vetting and static access control lists & Cryptographic identity and endorsement policies \\
\hline
Security (Tamper-resistance) & CVE history modifiable in database (with audit trail) & Immutable ledger with cryptographic hashing \\
\hline
CNA Revocation or Onboarding & Manual, out-of-band approval & Governed by blockchain-based access control and voting mechanisms \\
\hline
Public Auditability & Limited to published CVEs only & Full audit trail of drafts, updates, and embargoed states \\
\hline
\end{tabular}
\caption{Comparison of Current Centralized CVE Model vs Proposed Blockchain-based Model}
\label{tab:comparison}
\end{table}

\section{Related Work}
The existing CVE system, managed by the MITRE Corporation and overseen by the U.S. Department of Homeland Security, operates as a centralized infrastructure for collecting, validating, and publishing vulnerability disclosures. CNAs submit vulnerability reports to MITRE, which assigns CVE identifiers and maintains the public database \cite{third}. This process depends on a trusted third party (MITRE) and lacks decentralized oversight or cryptographic guarantees. As such, any disruption in MITRE's operations or changes in funding could jeopardize the reliability and accessibility of the CVE system.

Ruoti et al. \cite{fourth} provided a critical evaluation of blockchain's applicability across domains and emphasized that use cases like decentralized vulnerability disclosure can benefit from its properties of tamper-evidence and public verifiability. Similarly, Pilkington \cite{fifth} and Zheng et al. \cite{sixth} explored blockchain's potential in digital governance and transparency, laying the groundwork for applying it to cybersecurity infrastructure.

The classification of blockchain architectures into permissioned and permissionless types plays a key role in system design. Nakamoto \cite{seventh} and Narayanan et al. \cite{eighth} describe permissionless blockchains like Bitcoin, which allow any participant to read from and write to the ledger. These systems are fully decentralized but may suffer from scalability and governance issues. In contrast, permissioned blockchains, as detailed in Hyperledger Fabric's design by Androulaki et al. \cite{nineth}, limit write access to known participants and are better suited for enterprise and compliance-driven environments. Corda \cite{tenth} further extends this model with enhanced privacy and identity-based controls.

Szabo's foundational work \cite{eleventh} on smart contracts supports the idea of embedding enforcement logic directly into the blockchain, which is essential for features like embargoed disclosures and autonomous CNA governance. Tapscott and Tapscott \cite{twelveth} and Hegadekatti et al. \cite{thirteenth} have also discussed how blockchain can resolve trust issues in centralized institutions, reinforcing the case for decentralizing CVE management.

To our knowledge, no fully functional decentralized CVE publication system has yet been deployed. However, related research and proposals—such as Cosmos \cite{fourteenth} for interoperable blockchain networks or ZKP-based architectures discussed by Boneh and Shoup \cite{fifteenth}—demonstrate the feasibility of building scalable, privacy-preserving platforms for public interest applications.

\section{Future Work}
\par To further enhance privacy and scalability, we propose integrating Zero-Knowledge Proofs (ZKPs) to allow confidential submission validation without revealing sensitive details during the embargo period. For example, a ZKP can prove that a CVE follows schema and embargo constraints without disclosing its content until the embargo expires. Additionally, we aim to:

\begin{enumerate}
    \item Implement DAO-based governance to automate CNA onboarding/removal
    \item Add compatibility with existing CVE APIs to facilitate integration
    \item Explore IPFS or Filecoin for off-chain storage of large vulnerability reports
    \item Evaluate interoperability with other cybersecurity ledgers (e.g., threat intelligence)
\end{enumerate}

\section{Conclusion}
\par In this paper, we proposed a decentralized architecture for CVE publication based on a public permissioned blockchain. By leveraging Hyperledger Fabric, the system enforces strict authentication of CVE Numbering Authorities (CNAs), ensures public verifiability, and eliminates single points of failure. Smart contracts govern submission policies, embargo management, and revocation procedures. The proposed design addresses key limitations of the current centralized model governed by MITRE and establishes a foundation for secure, transparent, and scalable vulnerability disclosure.

While the current implementation relies on traditional certificate authorities (CAs) to authenticate CNAs, future work may incorporate decentralized identity (DID) frameworks, which enable entities to prove their identities using verifiable credentials without depending on a centralized root of trust. This shift would enhance the system’s resilience and align with the broader principles of decentralization. Additional future directions include the integration of zero-knowledge proofs and more robust decentralized governance mechanisms to further strengthen trust and automation. Beyond CVE management, the proposed architecture offers a versatile blueprint for applications in software supply chain transparency and cyber threat intelligence.

\bibliographystyle{unsrtnat}


\end{document}